\documentclass[letterpaper,prl,twocolumn,amsmath,amssymb,groupedaddress,bibliography]{revtex4-1}
\usepackage[utf8]{inputenc}
\usepackage[draft]{hyperref}
\usepackage{graphicx}
\usepackage{dcolumn}
\usepackage{bm}
\usepackage{amssymb}
\usepackage{physics}
\usepackage{color}
\usepackage{soul}
\usepackage{epstopdf}
\usepackage{float}

\begin{document}

\title {Thermal noise reduction in soliton microcombs via laser
self-cooling}
\author{Fuchuan Lei$^{1,\dag}$}
\email{fuchuan@chalmers.se} 
\author{Zhichao Ye$^{1,\dag}$}
\author{ Victor Torres-Company$^{1}$}

\affiliation{$^1$Department of Microtechnology and Nanoscience, Chalmers University of Technology SE-41296 Gothenburg, Sweden\\}

\date{\today}

\begin{abstract}
Thermal noise usually dominates the low-frequency region of the optical phase noise of soliton microcombs, leading to decoherence and limiting many aspects of applications. In this work, we demonstrate a simple and reliable way to mitigate this noise by laser cooling with the pump laser. The key is rendering the pump laser to simultaneously excite two neighboring cavity modes from different families that are respectively red and blue detuned, one for soliton generation and the other one for laser cooling. 
\end{abstract}

\pacs{Valid PACS appear here}
\maketitle


Microcombs are formed by pumping nonlinear microresonators with external laser sources. Microcombs have greatly endowed the family of frequency combs with new characteristics and functionalities \cite{del2007optical,diddams2020optical}. In contrast to traditional solid-state and fiber mode-locked laser frequency combs, the repetition frequency of microcombs can be easily set within a few  GHz to THz  through design of the dimension of the host microresonators. An effective strategy to achieve low-noise coherent microcombs is to generate dissipative solitons \cite{herr2014temporal,kippenberg2018dissipative}. 
 By proper dispersion engineering, octave-spanning soliton frequency combs are achievable, as required for self-referencing \cite{pfeiffer2017octave,li2017stably,weng2021directly}. With the potential of miniaturization and full integration, soliton microcombs have been proposed for many applications such as  spectroscopy \cite{suh2016microresonator}, lidar \cite{suh2018soliton,trocha2018ultrafast}, optical communications \cite{marin2017microresonator,fulop2018high}
 and astronomical applications 
\cite{obrzud2019microphotonic} among others. 

For all of the above-mentioned applications, the frequency or phase stability of
 comb lines is critical.
 According to elastic-tape model \cite{telle2002kerr}, recently adapted to the soliton microcombs \cite{lei2021fundamental}, the comb line frequency instability can be ascribed to the fluctuation of the pump and repetition frequencies. A frequency-stabilized and narrow-linewidth pump laser is  needed to minimize the soliton timing jitter \cite{lei2021fundamental}. In a soliton microcomb, the repetition frequency is not solely determined by the  microresonator free spectral range (FSR), but also coupled with the detuning through Raman self-frequency shift \cite{karpov2016raman} or spectral recoil
from dispersive waves or mode crossings \cite{brasch2016photonic, yi2017single}.
In addition, both the FSR and the cavity resonant frequency (therefore pump-resonator detuning) are functions of temperature due to thermo-optic and thermo-elastic effects, while the former is usually stronger \cite{gorodetsky2004fundamental}. Hence, the repetition frequency is influenced by thermal noise, specially the thermorefractive noise \cite{huang2019thermorefractive,drake2020thermal,panuski2020fundamental}. This noise source is particularly strong for small microresonators, as the temperature fluctuation is inversely proportional to the cavity-mode volume \cite{gorodetsky2004fundamental,huang2019thermorefractive}.

 Thermal noise induces soliton microcomb decoherence \cite{drake2020thermal}, which manifests in increased timing jitter or phase noise of the repetition frequency \cite{liu2020photonic,ye2021integrated} and broadened optical frequency lines
 \cite{nishimoto2020investigation,lei2021fundamental} . Recently, some methods have been proposed to mitigate the impact of thermal noise on soliton microcombs such as reducing the operation temperature \cite{moille2019kerr}, dispersion engineering \cite{stone2020harnessing}, and laser cooling \cite{drake2020thermal}. The latter technique makes use of an auxiliary laser that is placed to the blue side of a resonance far away from the pump mode to enable a thermal locking \cite{carmon2004dynamical}.  In this way, the inherent frequency jitter of the longitudinal cavity modes follow the (smaller) frequency drift of the auxiliary laser, and the pump laser "sees" a cavity as if it were cooled at cryogenic temperatures. To optimize the cooling effect,    both the cooling laser and pump laser should be well stabilized because their relative frequency drifting will cause  pump-resonance detuning fluctuations. This can be circumvented by introducing a single sideband modulation to the pump laser \cite{kuse2021phase} or pump-induced stimulated Brillouin scattering \cite{do2021self}. However, the sideband will be inevitably imprinted to the comb lines, thus limiting the comb purity unless a counter-propagation direction is utilized.


 \begin{figure}[t!]
\centering
\includegraphics[width=\linewidth]{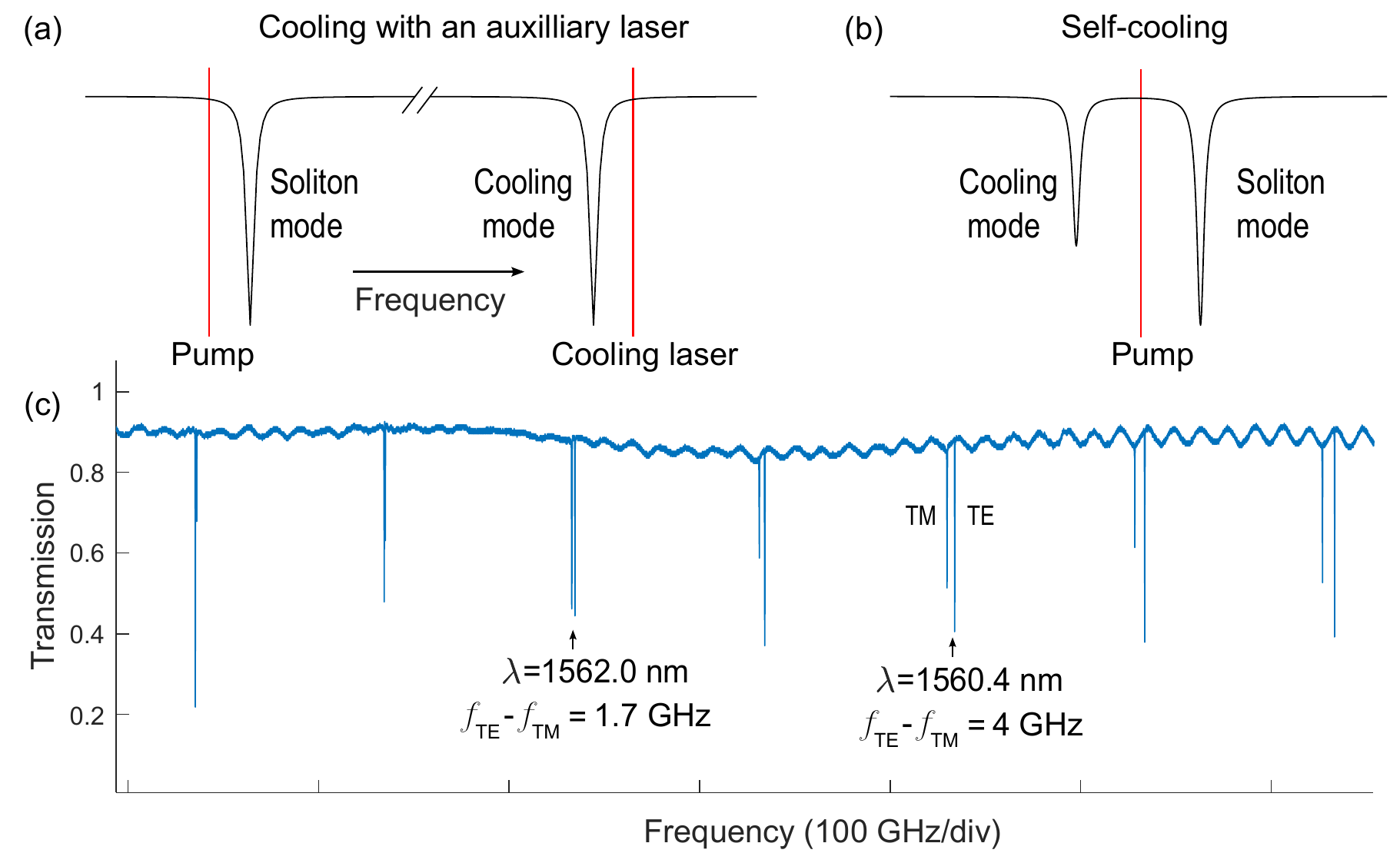}
\caption{Schematic of self-cooling of soliton microcombs. (a) Thermal noise mitigation through laser cooling. To obtain cooling effect, an auxiliary laser is applied to the blue side of another cavity mode in the same microresonator. (b) Cooling with the pump laser can be achieved if the pump is coupled simultaneously into one mode (red-detuned) for soliton generation and another mode (blue detuned) for cooling. 
(c) The transmission spectrum of the microresonator used in this experiment. To verify the effect of self cooling, we pump at two different wavelengths, where the detuning between soliton mode (TE) and cooling mode (TM) are different. }
\label{fig1}
\end{figure}

\begin{figure*}[t]
\centering
\includegraphics[width=0.95\linewidth]{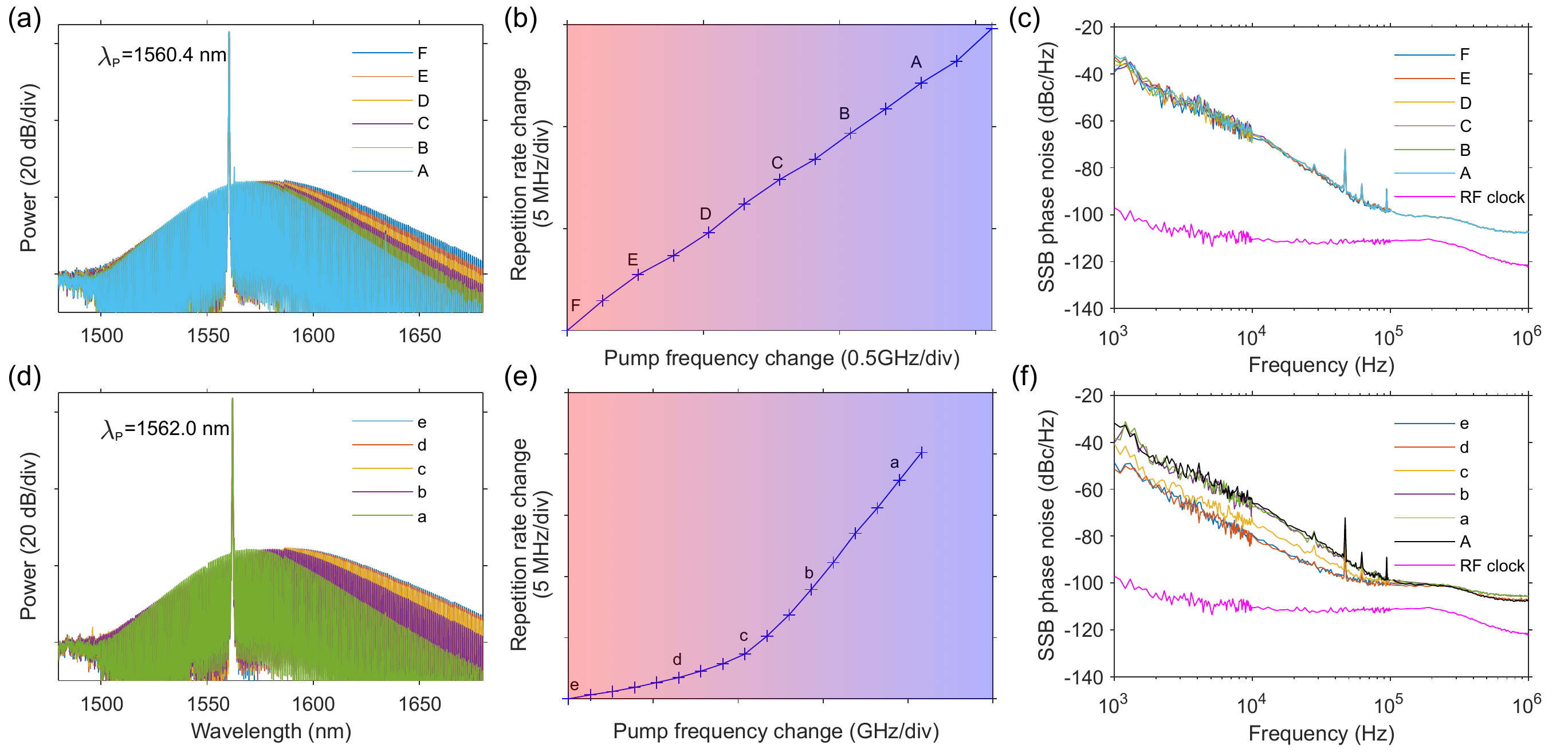}
\caption{Comparative study of the self-cooling of soliton microcombs. The top panel depicts the case without self-cooling effect, where the pump only excites a TE mode, (a)-(c) stand for the optical spectra, the repetition frequency and the singe side band (SSB) phase noise of the downconverted repetition frequency of the soliton microcombs at different pump detunings, respectively. (d)-(f) same for the case with self-cooling, where the pump is coupled to TE and TM modes simultaneously.}
\label{fig2}
\end{figure*}

Here we propose an alternative approach for laser cooling that avoids the problems of using an auxiliary laser.  Instead of involving an additional laser, we harness an auxiliary transverse cavity mode (cooling mode) in the vicinity and red side of the mode used for soliton generation (soliton mode).  A similar configuration has been demonstrated in optomechanical systems for simultaneously heating and cooling of mechanical motion \cite{tallur2012simultaneous}. 
The cooling mode should not be optically strongly coupled, but thermally coupled, to the soliton mode, i.e., their resonant frequencies should follow a similar shift with  temperature. In practice, the modes coming from the same microresonator are usually thermally correlated because they have nonzero spatial overlap.
The existence of a red-detuned neighboring mode could be beneficial for accessing the soliton state because it could help mitigate the temperature's sudden decrease in the soliton generation process   \cite{li2017stably,weng2021directly}, similar to the effect achieved with an auxiliary laser \cite{zhou2019soliton,zhang2019sub,geng2020enhancing}. However, its ability for thermal noise reduction has not been investigated. In this self-cooling configuration (Fig. \ref{fig1}), the pump laser plays a  twofold role: it cools the microresonator and generates a dissipative soliton microcomb. As a result, a self-cooled soliton state can be achieved. 



The self-cooling of soliton microcomb is experimentally studied with a silicon nitride ($\rm SiN$) microresonator. We generate a soliton at the fundamental TE modes while the cooling mode is selected from the fundamental TM modes.  The $\rm SiN$ microresonator is fabricated via subtractive processing  \cite{ye2019high}.  The width and height of the $\rm Si_3N_4$ waveguide are 1600 nm and 740 nm, respectively, which result in  the group velocity dispersion coefficients $\beta_{\rm 2}= -76 {\rm ps^2/km} $ for the fundamental TE mode and $\beta_{\rm 2}= -0.1 {\rm ps^2/km} $ for the fundamental TM mode, both evaluated at at 1562.0 nm. The free spectral ranges (FSRs) of TE and TM modes are 99.65 GHz and 98.47 GHz, and their averaged intrinsic (external) quality factors for the modes ranging from 1506 to 1630 nm are 7.9$\pm2.7$ (8.7$\pm2.0$) and 4.9$\pm3.6$ (6.3$\pm2.6$) million, respectively.


The FSR difference between TE and TM modes causes a walk off in the transmission spectrum of the microresonator, as shown in Fig. \ref{fig1}(c). To demonstrate the self-cooling effect, we performed a comparative study by pumping at either 1560.4 nm or 1562.0 nm. These two wavelengths correspond to different detunings between TE and TM modes. In principle, with properly engineering of the geometry of microresonators, it is always possible to attain close frequency degeneracy at the pump frequency for the two modes from different families involved in the soliton generation and cooling. The pump laser power is estimated to be $\sim$300 mW in the bus waveguide, which has identical cross-section geometry to the ring waveguide. With control of the polarization of the pump, $70\%$ ($30\%$) pump power is coupled to excite TE (TM) modes.  The soliton microcomb is initialized by scanning the  temperature of the microresonator with an integrated metallic heater  \cite{joshi2016thermally}. 

The results are presented in Fig. \ref{fig2}. When pumping at 1560.4 nm, to maintain the soliton state, the pump cannot excite TM mode because it is far detuned from the TE mode ($\sim$4 GHz). Therefore, the self-cooling effect cannot be achieved. In contrast,  both TE and TM modes can be excited simultaneously when we generate the soliton at 1562.0 nm as the cold cavity detuning between TE and TM mode is smaller ($\sim$1.7 GHz). The differences between the two cases can be revealed from the response of repetition frequency with respect to the change in pump frequency, as shown in Fig. \ref{fig2}(b) and (e). For both cases, the pump frequency is scanned across almost the full single-soliton existence range. 
The repetition frequency is measured via electro-optic downconversion \cite{del2012hybrid}, where a 25.1 GHz modulation is applied to the comb, resulting in a down-converted signal at $\sim$790 MHz. Due to the excitation of the TM mode, the repetition frequency does not longer evolve linearly with the pump frequency. As the pump laser is tuned towards the red, more pump power is coupled into the TM mode, leading to an increase in the resonator's temperature and  both TM and TE modes shifting towards longer wavelengths, hence the pump-resonance detuning is not changed as much as the case without the neighbouring TM mode being present. As a result, with the aid of the self-cooling action, the single soliton existence range in terms of pump frequency is significantly enlarged, which increases from $\sim$1.6 GHz to $\sim$4.2 GHz.  The soliton existence range can be further increased with proper parameters \cite{weng2021directly}. This is especially useful for some applications where the pump needs to be scanned or modulated, such as lidar \cite{liu2020monolithic}.

The phase noise of the repetition frequency is also indirectly obtained by measuring the  down-converted signal with a phase noise analyzer. This is feasible because the phase noise of the RF clock is extremely low, as shown in Fig. \ref{fig2}(c) and (f). Without self-cooling the phase noise is not related to the pump-resonance detuning.
However, at 1562.0 nm, the phase noise of the repetition frequency shows a considerable reduction as the pump frequency decreases. This indicates the cooling is strengthened with more power coupled to the TM mode. It should be noted that, to achieve an efficient self-cooling effect, the power coupled into the cooling mode should not exceed a threshold to avoid potential parametric oscillation and thermal instability.
However, the power used to excite the TM mode should not be too low either or the TM mode cannot be thermally locked to the pump. To illustrate
the latter point,  we tested pumping at 1562.0 nm with the input polarization completely parallel to the TE mode. In this case, the TM mode can also be excited due to the mode coupling with a strength of 2$\pi\times$150 MHz, however, this power coupled to TM mode is too weak for self-cooling, as the response of the repetition frequency and its phase noise show similarity to Fig. \ref{fig2}(b)-(c). This result also suggests the unavoidable scattering-induced optical coupling between TE and TM modes has a negligible impact here.  However, a stronger coupling could induce a strong dispersive wave \cite{yi2017single,yang2021dispersive}, and it may stabilize the soliton by trapping \cite{taheri2017optical}.

\begin{figure}[h]
\centering
\includegraphics[width=\linewidth]{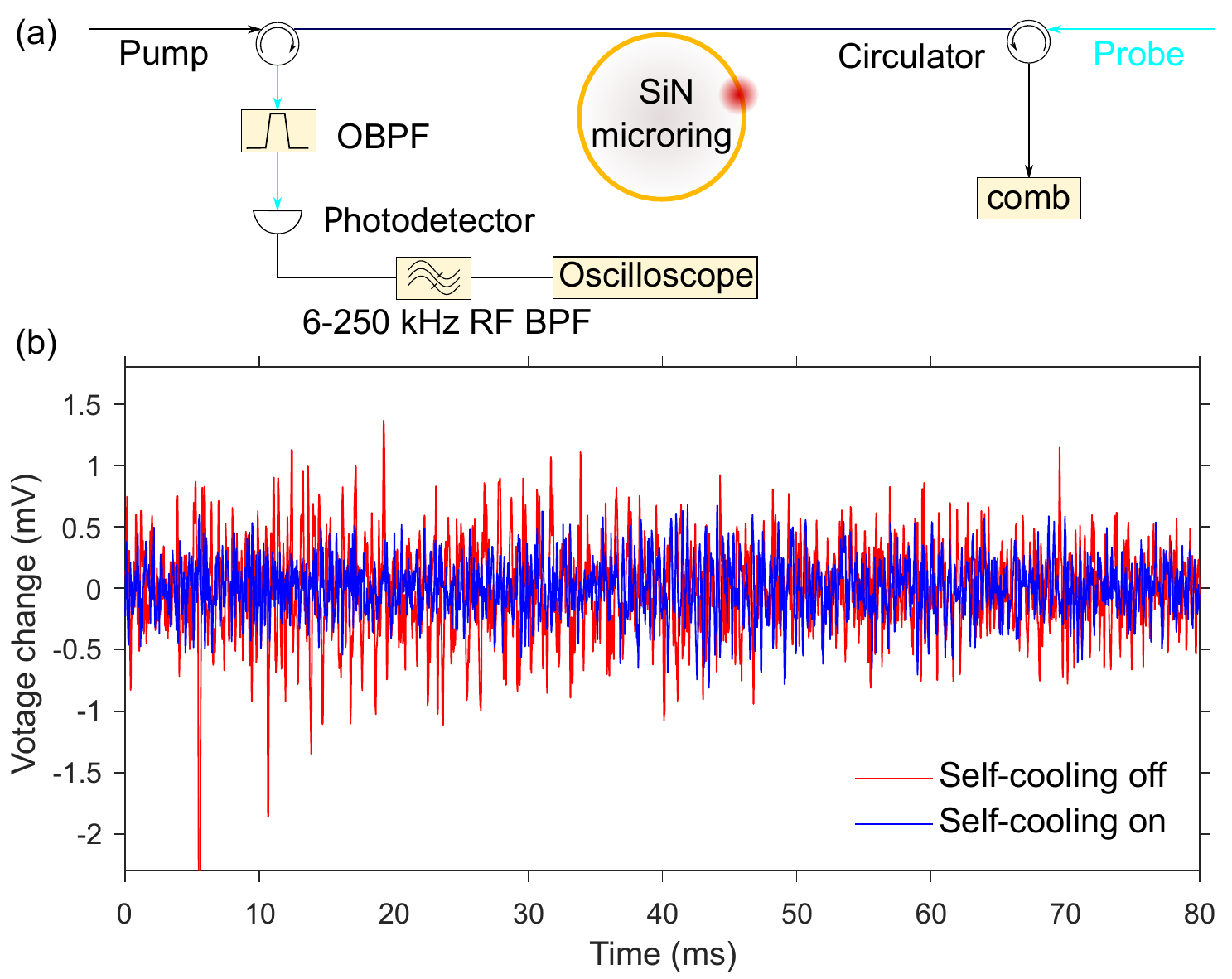}
\caption{Observation of thermal-noise reduction, which is manifested from the comparison of measured probe laser power (photodiode voltage) fluctuations for the cases with and without self-cooling.  An RF bandpass filter (BPF) from 6 kHz to 250 kHz is applied for better extracting of the thermal-noise induced signal. }
\label{fig3}
\end{figure}

To demonstrate the reduction of phase noise in repetition frequency is induced by the self-cooling action of the pump laser, we monitor the frequency fluctuations of a TE cavity longitudinal mode frequency when the pump is located at either 1560.4 nm (self-cooling off) or 1562.0 nm (self-cooling on). The results are presented in Fig. \ref{fig3}. The measurement system is similar to  \cite{drake2020thermal}, where a counterpropagating probe laser is coupled to a TE longitudinal mode at 1565.2 nm, away from the pump. Frequency-to-amplitude fluctuations are aided by the resonance lineshape and recorded in real time. Clearly, the self-cooling reduces significantly the frequency fluctuations of the cavity, in agreement with the reduction in repetition rate phase noise presented in Fig. \ref{fig2}.


\begin{figure}[h]
\centering
\includegraphics[width=0.9\linewidth]{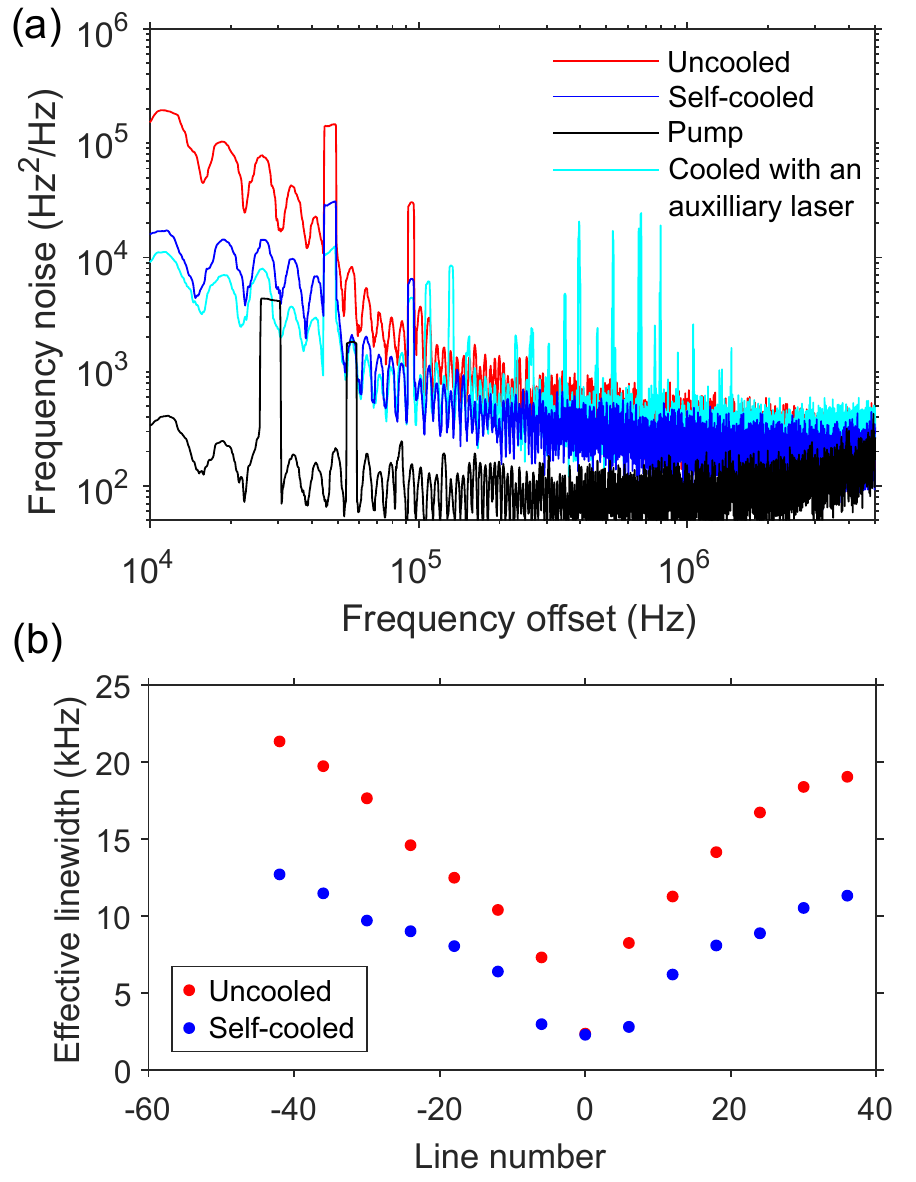}
\caption{Influence on the frequency noise and linewidth of comb lines. (a) The frequency noise power spectral density (PSD) of comb line $m=-42$ at different conditions. (b) Comparison of the comb lines' effective linewidth for soliton microcomb with and without self-cooling.  }
\label{fig4}
\end{figure}

According to the elastic tape model \cite{telle2002kerr}, the phase noise of repetition frequency is also transduced into optical phase/frequency noise in the comb lines.
This means that our self-cooling action should be encompassed by a reduction in optical linewidth of the microcomb lines. To demonstrate this point, comb line $m= -42$  (accounted from the pump at 0) at 1596 nm is chosen, as the effect of thermal noise is more 
evident for the comb lines far away from the pump. Its frequency noise is displayed in Fig. \ref{fig4}(a), which is measured with a self-heterodyne method \cite{lei2021fundamental}. It is clear that, compared to the soliton microcomb without self-cooling,  the low-frequency noise is dramatically reduced by more than 10 dB at 10 kHz offset.

Next, we compare the effectiveness of our self-cooling action to cooling with an auxiliary laser. The cooling laser (Santec TSL 710) is injected in opposite direction to the pump with the aid of circulators. The cooling laser is coupled into a TM mode and finally thermally locked at its blue-detuned side. The wavelength of the cooling laser is at 1565.2 nm. It is observed that, by optimizing the cooling laser's power and detuning, the low-frequency noise can be suppressed effectively, slightly better than the self-cooling. 
However, as anticipated, the frequency noise of cooling laser (the discrete spikes in the frequency noise power spectral density (PSD)) would be transferred into the comb lines.

Considering the effective linewidth is dominated by the low frequency region of the frequency noise,  it is meaningful to compare the effective linewidth $\Delta{\nu}_m^{\rm eff}$ for the soliton microcomb with and without self-cooling , which is calculated from the frequency noise PSD (${S_{\Delta\nu,m}(f)}$) through
  $ \int_{\Delta{\nu}_m^{\rm eff}}^{\infty} {S_{\Delta\nu,m}(f)}/{f^2} df={1}/{\pi}$ \cite{hjelme1991semiconductor}.
As demonstrated in Fig. \ref{fig4}(b), we can see the effective linewidth of all comb lines are notably reduced with self cooling.

To conclude, we have demonstrated a novel and simple way to reduce thermal decoherence of soliton microcombs. Instead of using an auxiliarly laser for cooling the cavity, we exploit the crossing with an auxiliary cavity mode.  A single pump laser can simultaneously generate a cavity soliton in the fundamental mode and leverage the thermal correlation with a different transverse cavity mode to enable thermal locking and laser cooling. This self-cooling action effectively damps thermorefractive noise and results into enhanced repetition rate stability and timing jitter, and a decrease in the effective linewidth of the soliton microcomb lines. 

\section{Acknowledgments} 
F.L and Z.Y contributed equally to this work. The devices demonstrated in this work were fabricated at Myfab Chalmers. This work is supported by European Research Council (CoG GA 771410); Vetenskapsrådet (2015-00535, 2020-00453). The authors declare no conflicts of interest.

\bibliography{ref}

\end{document}